# Measurement of the Tau Lifetime at SLD*

The SLD Collaboration

*Stanford Linear Accelerator Center*
*Stanford, CA 94309*

**Abstract**

A measurement of the lifetime of the tau lepton has been made using a sample of 1671 $Z^0 \rightarrow \tau^+\tau^-$ decays collected by the SLD detector at SLC. The measurement benefits from the small and stable collision region at SLC and the precision pixel vertex detector of SLD. Three analysis techniques have been used: decay length, impact parameter, and impact parameter difference methods. The combined result is $\tau_\tau = 297 \pm 9(stat.) \pm 5(syst.)$ fs.

*Work supported by U.S. Department of Energy contract DE-AC03-76SF00515.

# 1 Introduction

Measurements of tau lepton decays provide a unique test of lepton universality. Assuming the tau to have the same coupling to the W-boson as the muon and electron, the tau mass $m_\tau$, the tau lifetime $\tau_\tau$, and the $\tau \rightarrow e\nu_e\nu_\tau$ branching fraction $B_e$ are related as follows:

$$\tau_\tau = \tau_\mu \left(\frac{m_\mu}{m_\tau}\right)^5 B_e,$$

where $m_\mu$ and $\tau_\mu$ are the mass and lifetime of the muon. Currently the precision of this test is limited by measurements of the lifetime and the electronic branching ratio of the tau.

The Stanford Linear Collider (SLC) and the SLC Large Detector (SLD) provide an excellent facility for measuring the tau lifetime. Decays of tau pairs produced at the $Z^0$ mass peak are highly boosted and collimated, and are relatively easy to distinguish from other final states. In addition, the SLC is characterized by a very small and stable luminous region, while the SLD with its CCD pixel vertex detector has the capability of measuring three-dimensional space points with high resolution close to the interaction point. The combination of these features substantially reduces the uncertainty in the determination of both the production and decay points of the tau, allowing for precise measurements with a relatively small sample of events.

The results reported here are based on a sample of 1671 tau-pair candidates collected by SLD in 1992 and 1993. Three different techniques are used to determine the tau lifetime. In the first method, which uses tau-pair events in the 1 vs. 3 topology, the tau lifetime is extracted directly by measuring the decay length of the tau on the three-prong side of the event. The other two methods employ events in which both taus decay to a one-prong. In the impact parameter method, the tau lifetime is inferred from the distribution of impact parameters of the charged tracks, while in the impact parameter difference method, it is extracted from the correlation between the impact parameters and acoplanarity of the two tracks in the event. The decay length method gives a direct measurement of the tau lifetime, with relatively small backgrounds. The impact parameter and impact parameter difference techniques benefit from the large one-prong branching fraction of the tau, and are independent of the decay length method.

# 2 The SLD Detector

A detailed description of the SLD detector can be found in Ref. [1]. A subset of detector elements relevant to the analyses reported here are described briefly below. These include the vertex detector (VXD)[2], the central (or barrel) drift chamber (CDC)[3], and the lead/liquid-argon calorimeter (LAC)[4].

The VXD consists of 480 charge-coupled devices (CCDs) surrounding a 1 mm thick beryllium beam pipe with an inner radius of 25 mm. Each CCD is an array of $375 \times 578$ square pixels 22 $\mu$m on a side. The active material is a 20 $\mu$m thick epitaxial silicon layer on a 180 $\mu$m thick silicon substrate. The CCDs are mounted on 60 alumina boards 9.2 cm long, arranged in four concentric cylinders at radii ranging from 2.9 cm to 4.1 cm. The inner (outer) cylinder covers a range of polar angles defined by $|\cos\theta| < 0.85$ (0.75). The CCDs are arranged so that at least two hits are possible over the full azimuth within the polar angle acceptance. On average 2.3 CCDs are traversed by a track from the interaction point. The spatial resolution of the VXD is 6 $\mu$m transverse to the beam and 7 $\mu$m along the beam direction[6].

The barrel drift chamber in SLD is a cylinder 1.8 m long with an inner radius of 0.2 m and an outer radius of 1.0 m, filled with a $CO_2$-based gas mixture. There are 640 drift cells arranged in ten superlayers covering radii from 24 cm to 96 cm, where superlayers with axial wires alternate with pairs of stereo layers at angles of $\pm 43$ mrad. Each cell in a superlayer has eight sense wires spaced radially by 5 mm. Field-shaping wires in each cell provide the desired drift fields. Each sense wire provides a measurement of the drift distance with a spatial resolution averaging 70 $\mu$m over the entire drift cell. Tracks are reconstructed with high efficiency at polar angles in the range $|\cos\theta| < 0.85$.

The LAC consists of an assembly of rectangular lead plates separated by 2.75 mm gaps and mounted in large insulated vessels filled with liquid argon. A barrel section covers polar angles in the range $|\cos\theta| < 0.82$ and endcap sections complete the coverage down to $|\cos\theta| = 0.99$. The LAC is segmented in depth into an electromagnetic (EM) section and a hadronic section (HAD). The EM section is made with 2 mm thick lead plates for a total depth of 21 radiation lengths and 0.84 hadronic interaction lengths, while the HAD section has 6 mm thick plates for an additional depth of 2.0 interaction lengths. The lead plates are segmented and connected so as to form projective towers with an azimuthal segmentation of 33 mrad in the EM section and 66 mrad in the HAD section, and with comparable segmentation in polar angle. The energy resolution for electromagnetic showers has been measured to be about $\frac{\sigma_E}{E} = \frac{15\%}{\sqrt{E(\mathrm{GeV})}}$.

The SLD event trigger requires any of several combinations of tracking and energy-flow information from the detector elements. A subset of these have a relatively high efficiency for tau-pair events, in particular a requirement of two back-to-back tracks, or a single track plus a minimum energy deposition in the calorimeter.

## 2.1 Tracking Performance

Charged tracks found in the CDC are linked to clusters of pixels in the VXD by extrapolating each track and selecting the best set of associated clusters[6]. A set of clusters may not be shared by multiple tracks. The track parameters are then recalculated, accounting for multiple scattering. In tau-pair events, at least one VXD cluster is linked to a well-measured CDC track in 99% of the cases.

The measured impact parameter resolution transverse to the beam line for the combined tracking system is 11 $\mu$m for infinite momentum tracks and 76 $\mu$m at 1 GeV/c. Along the beam direction the resolution is 38 $\mu$m for infinite momentum and 80 $\mu$m at 1 GeV/c. The momentum resolution of the tracking system in the 0.6 T field of the SLD solenoid is $\left(\frac{\delta p_T}{p_T}\right)^2 = 0.01^2 + (0.0026\ p_T)^2$, where $p_T$ is the track transverse momentum in GeV/c.

## 2.2 Interaction Point Determination

The SLC collides bunches of electrons and positrons accelerated in the SLAC linac at a rate of 120 Hz. After colliding at energies of 45.6 GeV each, the bunches are extracted and dumped. For the 1993 run, the spatial extent of the bunches at the interaction point was typically $\approx 0.8$ $\mu$m vertically, $\approx 2.6$ $\mu$m horizontally, and $\approx 700$ $\mu$m longitudinally[5]. The transverse position of the SLC collision region was stable, with variations of typically 5-10 $\mu$m over time periods measured in hours.

The spatial location of the interaction point is determined accurately in the transverse plane using tracks from hadronic $Z^0$ decays [6]. A fit to a single point in the transverse plane is made using tracks from 30 successive events, in an iterative procedure. At each iteration, tracks that significantly reduce the quality of the fit are removed. Typically the fit uses about 300 tracks, and converges after a few iterations. The uncertainty in this determination is $7 \pm 2$ $\mu$m for the 1993 data and $9 \pm 2$ $\mu$m for the 1992 data. Non-Gaussian tails in the interaction point distribution may be represented conservatively by a second Gaussian with a standard deviation of 100 $\mu$m for 0.25% (0.5%) of the 1993 (1992) events.

# 3 Event Selection

The events used in the present analyses were selected from a sample of 2.20 pb$^{-1}$ collected at a center-of-mass energy of 91.2 GeV in 1992 and 1993. The selection of tau-pair candidates is based on the multiplicity, momentum and direction of tracks in the central drift chamber, and on properties of electromagnetic showers in the calorimeter. Tracks and energy clusters are required to meet certain criteria[7] to be used in the event selection.

## 3.1 Selection Criteria

Tau-pair candidates are required to have at least two but fewer than seven tracks. Each event is divided into hemispheres by the plane normal to the track with the highest momentum. Tracks in each hemisphere must fall within 15° of the net momentum vector in the hemisphere, and the jet invariant mass[8] in each hemisphere is required to be less than 2.3 GeV/c$^2$. Furthermore, the jet axes in the two hemispheres must be back-to-back within 20°. These criteria discriminate strongly against background from multi-hadron final states. The polar angle $\Theta_{miss}$ of the missing momentum[8] in each event is required to satisfy $|\cos\Theta_{miss}| < 0.88$ to discriminate against two-photon interactions and Bhabha events. Two-prong events are required to have a minimum acolinearity of 10 mrad. The scalar sum of the momenta of the two stiffest tracks in any event must be less than 65 GeV/c. These cuts primarily reject Bhabha events and muon-pair final states.

The total visible energy[8] in an event is required to be at least 12% of the center-of-mass energy ($E_{CM}$) to reject two-photon interactions. To discriminate against Bhabha events, the total energy deposited in the electromagnetic section of the calorimeter is required to be less than 0.45 $E_{CM}$, and the most energetic EM cluster must be less than 0.33 $E_{CM}$. In addition, the total calorimeter energy not included in identified jets[8] is required to be less than 5 GeV, and there must be fewer than six energy clusters not included in jets.

These criteria resulted in a sample of 1671 tau-pair candidates selected from the 1992 and 1993 data. Further requirements were imposed in each individual lifetime analysis.

## 3.2 Selection Efficiency and Backgrounds

The event selection efficiency and background contamination were estimated using Monte Carlo. The production of tau-pair events at the $Z^0$ resonance was simulated using the KORALZ 3.8[9] Monte Carlo generator. The same program was used to generate muon-pair events, while wide-angle Bhabha scattering, two-photon interactions, and $Z^0 \rightarrow hadrons$ final states were produced using the generators described in Refs. [10], [11], and [12], respectively. All these Monte Carlo data samples were subjected to the SLD detector simulation based on the GEANT 3.15[13] program and to the above event selection. The SLD trigger was also simulated in the Monte Carlo.

We determine a tau-pair event selection efficiency of $63.0 \pm 0.3\%$, where the largest loss is due to the solid angle coverage of the CDC. The trigger efficiency is found to be $95.1 \pm 0.3\%$, and the result is an overall selection efficiency of $59.9 \pm 0.4\%$. Based on these estimates, the measured integrated luminosity, and the $Z^0 \to \tau^+\tau^-$ branching fraction, we expect a final tau sample of $1714 \pm 50$ events, including backgrounds. This is consistent with the 1671 tau-pair candidate events selected in the data. From the fraction of Monte Carlo background events which survive the selection requirements, the purity of the tau sample is estimated to be $96.1 \pm 0.4\%$. The accuracy of these estimates was checked by comparing data and Monte Carlo distributions for various quantities used in the event selection.

### 3.3 Selection of Specific Topologies

**1-3 events:** The decay length method (DL) described below uses only events with a 1-3 topology. Any pair of oppositely charged tracks which are consistent with originating from a photon conversion [6] are excluded. An event is then required to have exactly three tracks which form a common vertex in one hemisphere, and a single charged track in the opposite hemisphere. Each of the three tracks must have at least 25 hits in the CDC and at least one hit in the VXD, and the $\chi^2$ per degree of freedom of the track fit must be less than 5 for two of the tracks and less than 15 for the third. The $\chi^2$ probability for the vertex fit must be at least 0.02%. This results in a final sample of 257 events.

**1-1 events:** For the two methods employing impact parameters, only events with a 1-1 topology are used (here photon conversions are also removed). For the direct impact parameter (IP) method, at least one track in each event must have: a momentum greater than 3 GeV/c and less than 40 GeV/c, at least 40 CDC hits and at least one hit in the VXD, a polar angle in the range $|\cos\theta| < 0.72$, a $\chi^2$ per degree of freedom for the track fit less than 5.0, and a distance of closest approach to the interaction point along the beam direction less than 2.5 mm. For the impact parameter difference (IPD) method, both tracks in each event must satisfy these criteria. For both methods the tracks must have opposite charge, each event must have a two-prong invariant mass of at least 8 GeV/c$^2$, the angle between the two tracks must be at least 2.8 rad, and the missing momentum in the event must satisfy $|\cos\Theta_{miss}| < 0.80$. These criteria result in a sample of 1556 tracks (from 912 events) for the IP method, and 642 events for the IPD method.

| Process | Background fraction (%) | | |
|---|---|---|---|
| | DL | IP | IPD |
| Multihadron | $0.3 \pm 0.1$ | | |
| Muon-pair | | $0.9 \pm 0.1$ | $0.2 \pm 0.1$ |
| Bhabha | | $0.6 \pm 0.1$ | $0.1 \pm 0.1$ |
| Two-photon | | $1.0 \pm 0.3$ | $0.5 \pm 0.3$ |
| Total | $0.3 \pm 0.1$ | $2.5 \pm 0.3$ | $0.8 \pm 0.3$ |

Table 1: Background fractions (in percent) in the final event samples used in the DL, IP, and IPD analyses.

The main contamination for 1-3 events is from multihadron events, whereas for events in the 1-1 topology the primary backgrounds are Bhabha scattering, two-photon interactions, and muon-pair final states. Table 1 summarizes the background fractions in the final event samples used in each of the three tau lifetime measurements.

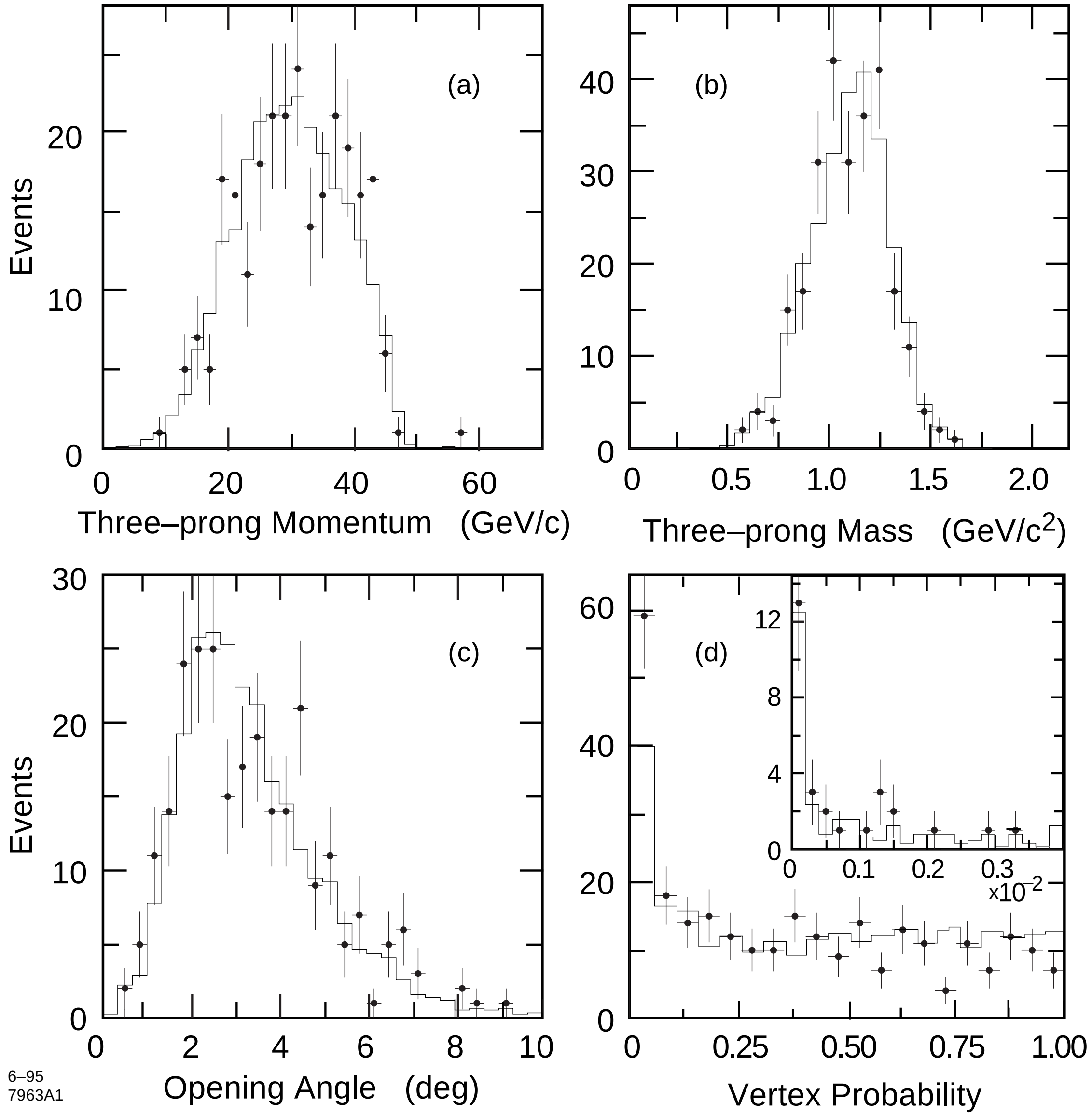

Figure 1: Distributions from the 1-3 sample: (a) three-prong momentum, (b) three-prong invariant mass, (c) largest opening angle between a track and the three-prong momentum direction, and (d) three-prong vertex fit probability. In (d) the inset shows an expanded view of the lowest 0.4% in probability.

A few track and event quantities are plotted in Fig. 1 and Fig. 2 for tau-pair events in the 1-3 and 1-1 topologies, respectively. In all plots, the dots represent data and the histograms Monte Carlo. The three-prong momentum and invariant mass for 1-3 events are shown in Fig. 1(a) and 1(b), respectively. Figure 1(c) shows the largest opening angle between the three-prong momentum direction and one of the three tracks, while in Fig. 1(d) is plotted the probability of the three-prong vertex fit. The distributions in Fig. 2 for 1-1 events represent (a) the track momentum, (b) the track fit $\chi^2$ per degree of freedom, (c) the invariant mass of the two tracks and (d) their acolinearity. Good agreement is seen in all the distributions, indicating sufficient accuracy in the Monte Carlo simulation of the data.

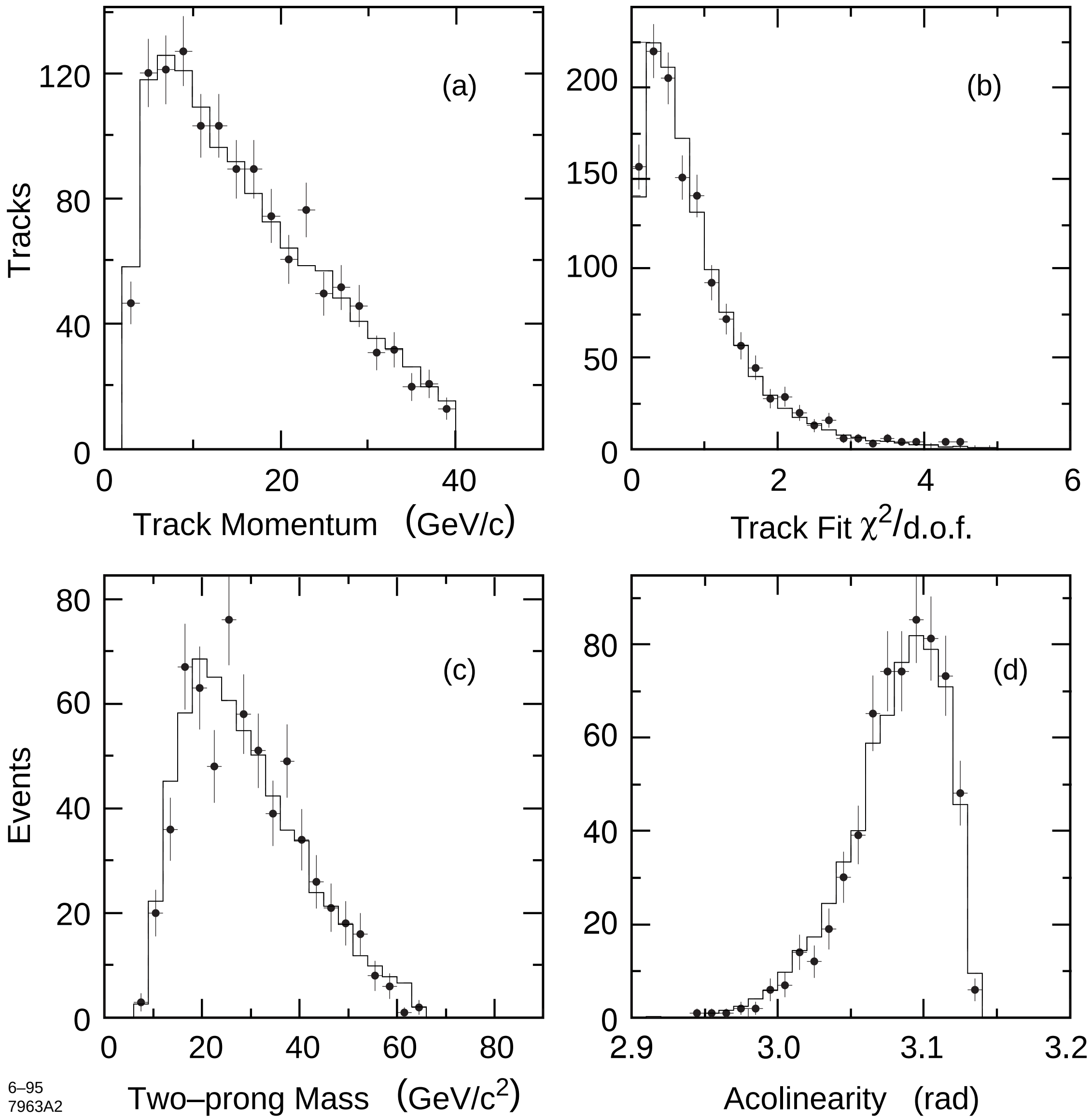

Figure 2: Distributions from the 1-1 sample: (a) track momentum, (b) track fit $\chi^2$/d.o.f., (c) two-prong invariant mass, and (d) two-prong acolinearity.

# 4 Lifetime Measurement

The three techniques used in this analysis are illustrated in Fig. 3. Figure 3(a) shows the principle of the decay length method. Using tau-pair events with a 1-3 topology, selected as described above, a vertex fit in three dimensions is performed on the three-prong side to determine the decay point of the tau. Because of the relatively large beam size in the longitudinal direction, a two-dimensional decay length is calculated for each event using the precisely measured transverse beam position and the projection of the three-prong vertex in the transverse plane. The result is then translated into a three-dimensional decay length using the event thrust axis.

Figure 3(b) illustrates the impact parameter method. In this technique, events in the 1-1 data sample are used. For each event, either one or two measurements are extracted according to whether one or both of the two charged tracks in the event satisfy the selection requirements described in section 3.3. The impact parameter $d$ of a daughter track is related to the transverse decay length $l_{xy}$ of the parent tau as follows (see Fig. 3(b)):

$$d = l_{xy} \sin\phi = l \sin\theta \sin\phi,$$

where $\phi$ is the angle between the track and the direction of the parent tau, and $\theta$ the polar angle of the tau direction.

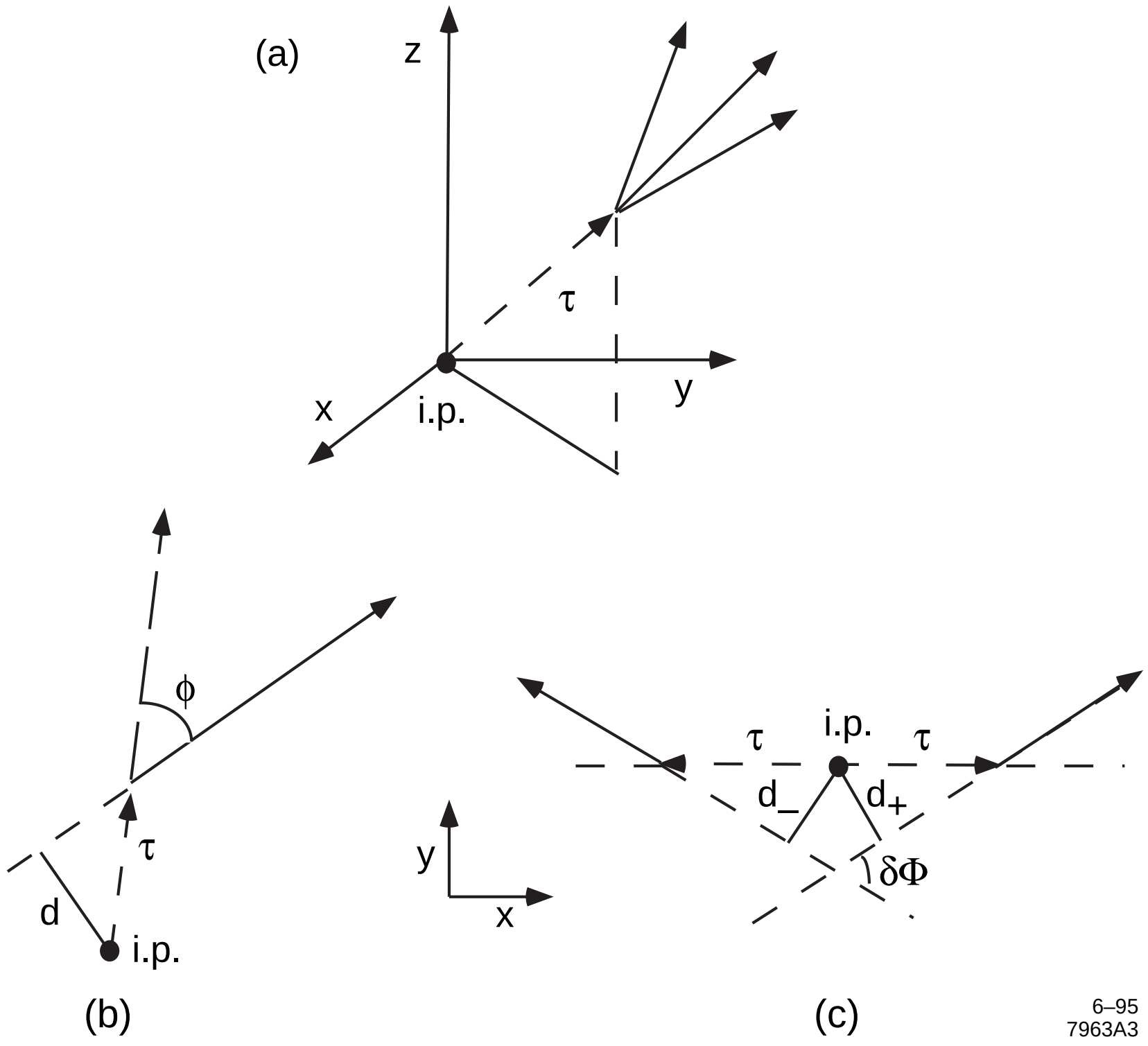

Figure 3: Drawings illustrating the three measurement techniques used: (a) decay length, (b) impact parameter, and (c) impact parameter difference. The x and y axes lie in the transverse plane.

The impact parameter difference technique [14] is illustrated in Fig. 3(c). Using events with a 1-1 topology, selected as described in section 3.3, this method exploits the correlation which exists between the difference of the impact parameters of the two tracks in the event and their acoplanarity. This is expressed by the following equation:

$$< d_+ - d_- >=< l_{xy} > \ \delta\phi =< l > \ \sin\theta \ \delta\phi,$$

where $d_+$ and $d_-$ are the impact parameters of the two tracks, $\delta\phi$ is the acoplanarity of the tracks, and $\theta$ is the polar angle of the tau direction.

Directly or indirectly, all three methods involve the measurement of an average decay length $< l >$. The tau lifetime $\tau_\tau$ is related to $< l >$ by: $< l >=< \beta\gamma > c\tau_\tau$, where $< \beta\gamma >$ is the average boost of the tau and is determined from Monte Carlo. In all three methods, the tau direction is taken to be the event thrust axis, which is determined on an event-by-event basis from charged tracks and isolated energy clusters in the calorimeter. Monte Carlo studies show this to be a very good approximation.

In the decay length (DL) method, the result is derived from an unbinned maximum likelihood fit to the decay length distribution using an analytical function given below. In both the impact parameter (IP) and impact parameter difference (IPD) techniques, we make use of a binned maximum likelihood fit using Monte Carlo to parameterize the experimental quantities that are used in the two measurements: impact parameter for the IP method, and impact parameter difference and acoplanarity for the IPD method.

## 4.1 Decay Length Method

The distribution of measured decay lengths for the 257 selected events is plotted in Fig. 4(a). The average decay length is extracted from a maximum likelihood fit using an exponential decay distribution function convoluted with a Gaussian resolution function:

$$P(l_i, \sigma_i; l_0, s_0) = \frac{1}{l_i s_0 \sigma_i \sqrt{2\pi}} \int \exp\left(-\frac{x}{l_0}\right) \exp\left(-\frac{(x - l_i)^2}{2 s_0^2 \sigma_i^2}\right) dx,$$

where the $l_i$ are the measured decay lengths, $l_0$ is the parent decay length, and $s_0$ is a scale factor on the calculated decay length errors $\sigma_i$. The $\sigma_i$ are determined on an event by event basis from a projection of the three-dimensional vertex error matrix along the event thrust axis, combined with a small contribution from the uncertainty in the transverse beam position. The mean value of the decay length error for these events is calculated to be $475 \pm 9$ $\mu$m. From the fit, the average decay length is found to be $2.19 \pm 0.14$ mm, with a scale factor $s_0 = 1.31 \pm 0.11$. The solid curve in Fig. 4(a) represents the fit function with these values, normalized to the data points. As a measure of the goodness of the fit, the $\chi^2$ per degree of freedom for the normalization is 0.8.

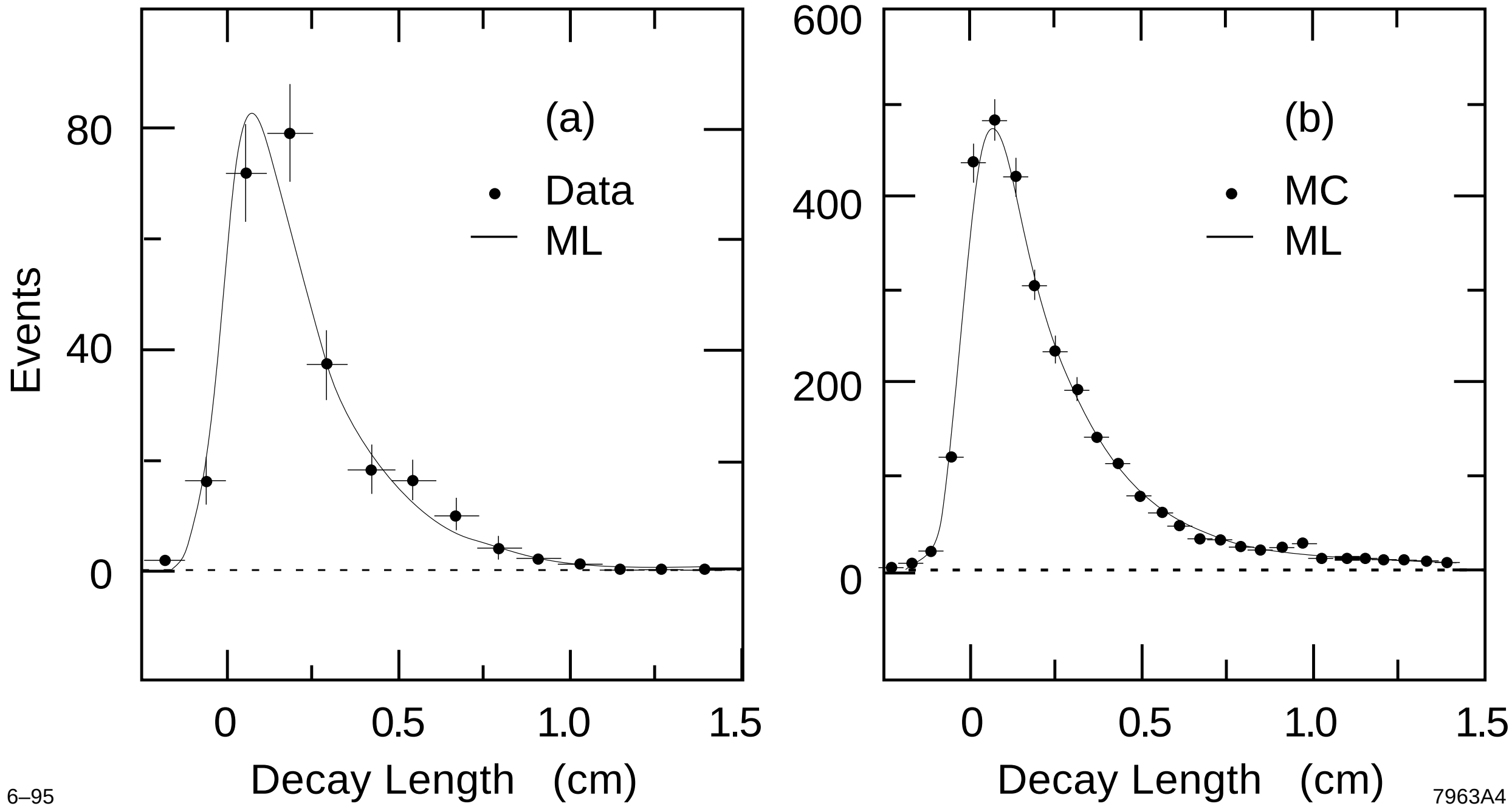

Figure 4: Three-prong decay length distribution for the final event samples in (a) the data and (b) the Monte Carlo. The solid curve in the two figures corresponds to the maximum likelihood fit described in the text.

Figure 4(b) shows the decay length distribution for 2846 Monte Carlo tau-pair events generated with a mean lifetime of 305 fs and passing the same event selection criteria as the data. The fit yields an average decay length $l_0 = 2.331 \pm 0.044$ mm, corresponding to a mean lifetime of $305.6 \pm 5.8$ fs, in good agreement with the generated value. The average decay length error for these events is 460 $\mu$m, and the fitted scale factor is $s_0 = 0.96 \pm 0.05$. Since the maximum likelihood fit to the Monte Carlo events returns the generated lifetime, no bias is attributed to the method and no associated systematic error is assigned.

A Monte Carlo study using a large number of event samples showed that the fitted value of the scale factor $s_0$ from the data is consistent with a statistical fluctuation from unity, as the distribution of $s_0$ is skewed toward larger values. As can be seen in Fig. 4, $l_0$ is sensitive primarily to events with relatively large decay length, while $s_0$ is determined mainly from decay lengths near zero, so that the lifetime result is fairly insensitive to the value of the scale factor. This was checked in the data by repeating the maximum likelihood fit with a fixed scale factor of unity. From the corresponding change in the fitted average decay length, a conservative systematic error of 0.8% in the lifetime is assigned.

The effect of the track quality requirements and the cut on the vertex fit probability was studied in the data. From the changes in the result when these quantities were varied, we derive a systematic error of 1.0%. The sensitivity of the measurement to the event selection requirements was studied in the Monte Carlo, and the systematic uncertainty was found to be negligible. As seen in Table 1, the only significant background found to survive all the analysis cuts comes from multihadron final states at a level of about 0.3% of the 1-3 sample. It was determined that the average lifetime of such events is consistent with zero, and that including this background reduces the fitted decay length by 0.3%. Due to the uncertainty in the modeling of low-multiplicity hadronic events, a conservative systematic error equal to the full amount of this correction is assigned.

| Systematic effects | Error (%) |
|---|---|
| Decay length resolution | 0.8 |
| Track and vertex quality cuts | 1.0 |
| Backgrounds | 0.3 |
| Initial and final state radiation | 0.3 |
| Beam energy and energy spread | 0.3 |
| Total | 1.4 |

Table 2: Systematic errors for decay length method.

To study the effect of a possible misalignment between the VXD and CDC, the data were divided into four samples by azimuthal quadrants of the detector where decay vertices were found. Lifetimes determined separately for the four samples were the same within statistical errors, and the associated systematic uncertainty is estimated to be negligible. The effect of non-Gaussian tails in the distribution of beam positions has been estimated to be negligible in this measurement. The two-dimensional projection of the decay length using the thrust axis was checked by using instead the net momentum of the three-prong to define the tau direction, and results were found to be consistent. The uncertainty in the calculation of initial- and final-state radiation in the Monte Carlo was estimated to contribute a systematic error of 0.3% in the average boost of the taus. Finally, the effect of uncertainty in the SLC beam energy and energy spread was studied by varying these quantities in the Monte Carlo for tau events and observing the change in the average tau momentum. A systematic uncertainty in the lifetime of 0.3% is estimated from this source.

The systematic effects discussed above are summarized in Table 2. From the Monte Carlo, we find for the average boost of the tau in these events: $<\beta\gamma>= 25.44 \pm 0.01$. Applying this factor, and including background correction and systematic errors, the decay length method yields a tau lifetime

$$\tau_\tau = 288 \pm 18 \pm 4 \text{ fs}.$$

This measurement was checked by an independent analysis in which the decay length was computed in three dimensions and considerable attention was paid to reducing the errors on the longitudinal position of the interaction point[15]. A consistent result was obtained.

## 4.2 Direct Impact Parameter Method

As described in section 3.3, the event selection yielded a total of 1556 tracks for this lifetime determination. The distribution of impact parameters measured for these tracks is shown in Fig. 5. The impact parameter is assigned a positive (negative) value if the extrapolated track crosses the event thrust axis before (after) its point of closest approach to the interaction point. Negative impact parameters result from finite tracking errors and uncertainties in the beam position determination.

To extract a lifetime from the impact parameter distribution, a binned maximum likelihood fit to the data is performed. The fit function is represented by the impact parameter distribution from the Monte Carlo, corrected for background and normalized to the number of events in the data. The likelihood probability is expressed as follows:

$$\ln L = \sum_{i=1}^{N} Y_i \ln f_i(\theta),$$

where N is the number of bins in the distribution used in the fit, $Y_i$ is the number of entries in the $i^{th}$ data bin, and $f_i(\theta)$ is the normalized content of the $i^{th}$ bin in the Monte Carlo. The bin width chosen is 10 $\mu$m, except that in the tails of the Monte Carlo distribution the bins are widened as required to include at least 10 events. The data are then binned the same as the Monte Carlo.

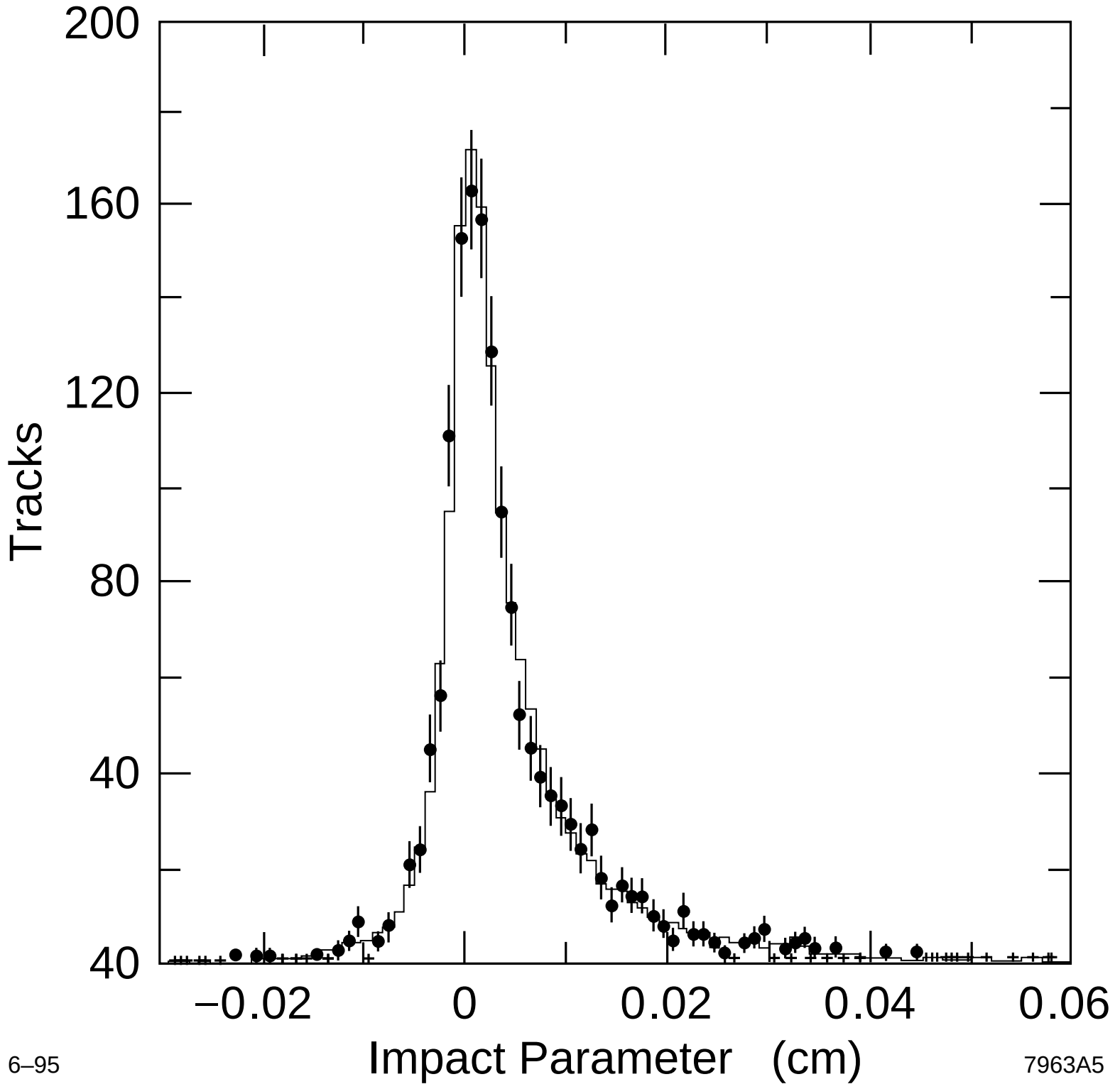

Figure 5: Impact parameter distribution for data (data points) and Monte Carlo (histogram).

A single Monte Carlo sample was generated with an input lifetime of $\tau_0 = 305$ fs. A weighting technique is then used to generate the impact parameter distributions corresponding to alternative lifetimes. A track originating from a tau decay with proper time $t$ is taken to have been produced from a sample with different lifetime $\tau$ by weighting its assigned probability with the following ratio:

$$weight = \frac{\tau_0}{\tau} \frac{\exp(-t/\tau)}{\exp(-t/\tau_0)}.$$

Weighted impact parameter distributions are formed with seven different lifetimes in the range $0.7\tau_0 < \tau < 1.3\tau_0$. These distributions are corrected for background by adding the normalized impact parameter distribution for background tracks passing all cuts. A likelihood probability is calculated for each distribution, and a fourth-order-polynomial fit is performed to the seven values as a function of lifetime. The best fit lifetime is taken to be the value corresponding to the maximum of this curve, with a statistical error assigned by taking the values where the likelihood has decreased by 0.5 from the maximum. The result is

$$\tau_\tau = 302 \pm 12 \text{ fs}.$$

The impact parameter distribution corresponding to the best fit lifetime in the Monte Carlo, including the background correction, is shown as the solid histogram in Fig. 5. The agreement between data and Monte Carlo is very good, and the comparison between the two distributions gives a $\chi^2$ of 0.9 per degree of freedom.

This analysis was checked for systematic bias using several different Monte Carlo samples of 1000 tracks each in place of the data. In every case the generated lifetime was reproduced within the statistical errors. A similar check was made with several groups of Monte Carlo events using the true generated tau direction in place of the event thrust axis, and results were consistent within statistics. The statistical error obtained from the likelihood fit was found to be consistent with the fluctuations in the results from the different Monte Carlo samples.

The sensitivity of the fitting technique to bin size and Monte Carlo statistics was studied extensively. The analysis was repeated with nominal bin sizes a factor of two larger and smaller than the chosen 10 $\mu$m, and the minimum number of events required in a Monte Carlo bin was also varied. The corresponding variations in the result lead to a systematic uncertainty of 0.3%. A study using several Monte Carlo samples of different sizes lead to an estimated systematic error of 0.9% due to finite Monte Carlo statistics. The range of impact parameters covered by the fit was varied, equivalent to a symmetric trimming of the sample from zero to 5%, both for the data and Monte Carlo. The results fluctuate within statistical errors, and no associated systematic error is assigned.

| Systematic effects | Error (%) |
|---|---|
| Binning | 0.3 |
| Monte Carlo statistics | 0.9 |
| Track quality cuts | 1.0 |
| Backgrounds | 0.1 |
| Beam position | 0.4 |
| Tau branching ratios | 0.2 |
| Initial and final state radiation | 0.3 |
| Beam energy and energy spread | 0.3 |
| Total | 1.5 |

Table 3: Systematic errors for the impact parameter method.

All track quality cuts were varied in order to study their effect on the measurement. From the observed changes in the measured lifetime, we assign a conservative systematic error of 1.0%. The uncertainty in the background correction was checked by varying the background fraction in the Monte Carlo, and a corresponding systematic error of 0.1% is assigned.

The possibility of detector misalignment was checked as described above for the decay length method, and no statistically significant effect was observed. The effect of non-Gaussian tails in the beam position determination was studied in the Monte Carlo by introducing an additional smearing in the beam position in a fraction of events. The change in the measured lifetime was small and a conservative systematic error of 0.4% is assigned. We have also investigated the effect of uncertainties in the tau one-prong branching ratios used in the Monte Carlo. We estimate a systematic error of 0.2% from this source. Finally, as mentioned above for the decay length method, we assign a systematic error of 0.3% due to initial- and final-state radiation, and another 0.3% due to the uncertainty in the beam energy and beam-energy spread.

A summary of the systematic errors is given in Table 3. Including systematic errors, the direct impact parameter method yields a tau lifetime

$$\tau_\tau = 302 \pm 12 \pm \ 5 \text{ fs}.$$

## 4.3 Impact Parameter Difference Method

As described in section 3.3, the event selection yielded a sample of 642 events for this analysis. The impact parameter difference and acoplanarity distributions for these events are plotted in Fig. 6(a) and 6(b), respectively[16]. The data (represented by the dots) and the Monte Carlo (by the histograms) are seen to be in good agreement. In Fig. 7(a), the scatter plot of impact parameter difference versus acoplanarity for the data is shown. A clear correlation is evident, and the tau lifetime can be extracted from the slope of the distribution. This is even better illustrated in Fig. 7(b) where each data point represents the mean impact parameter difference $< d_+ - d_- >$ over a given bin in $\sin\theta\ \delta\phi$.

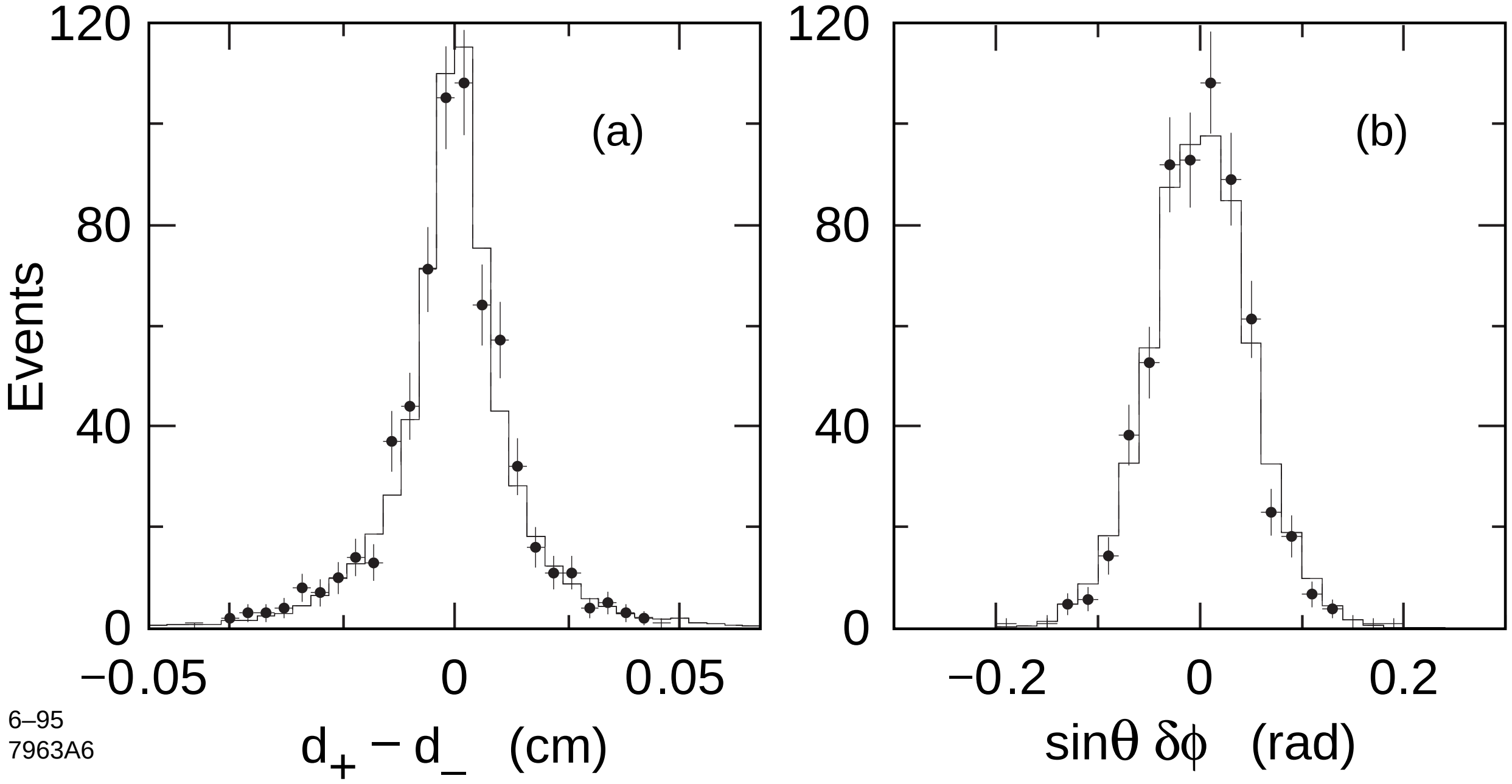

Figure 6: (a) Impact parameter difference and (b) acoplanarity distributions for the final event sample in the IPD method. The dots represent the data and the histograms the Monte Carlo.

A straightforward way to extract the tau lifetime is to fit the distribution in Fig. 7(b) to a straight line over a range in acoplanarity where the correlation is linear, together with a trimming procedure in order to remove from the fit tails that are not well modeled. In our case, we have chosen an alternative method based on a binned maximum likelihood method. This is motivated by the fact that in any acoplanarity bin, the impact parameter difference distribution is asymmetric and non-Gaussian (it is exponentially distributed) and any truncation would result in a bias in the lifetime.

The maximum likelihood technique used here is similar to that described in section 4.2 for the single-impact parameter method; the only difference is that the data are described in two-dimensional bins in impact parameter difference and acoplanarity. Just as in the impact parameter method, a single Monte Carlo sample was generated, and several samples with different lifetimes were simulated using a weighting technique with a weight defined by:

$$weight = \frac{\tau_0^2}{\tau^2} \frac{\exp\left(-\frac{(t_1+t_2)}{\tau}\right)}{\exp\left(-\frac{(t_1+t_2)}{\tau_0}\right)},$$

where $\tau_0$ is mean lifetime in the original Monte Carlo sample, $\tau$ the alternative lifetime, and $t_1$ and $t_2$ the decay times of the two taus in the event.

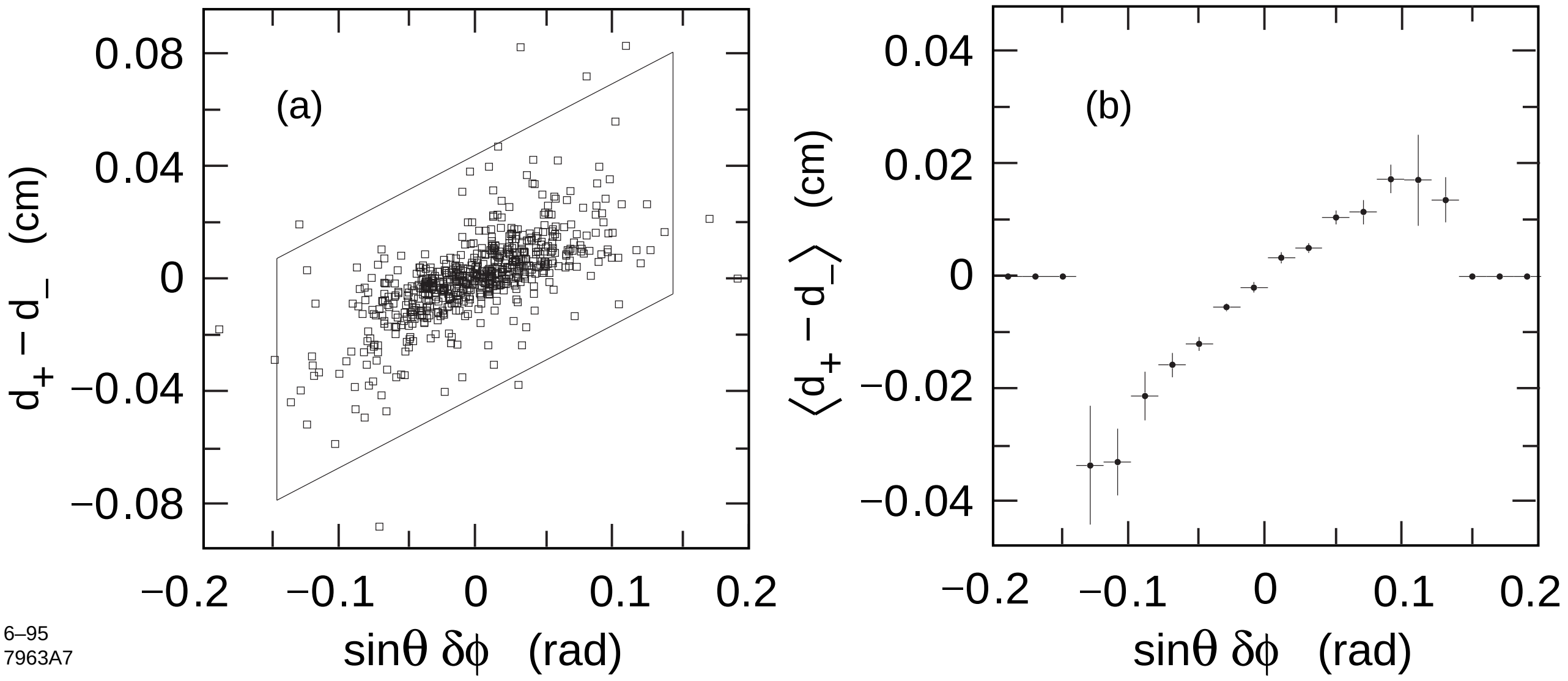

Figure 7: Impact parameter difference vs. acoplanarity scatter plot for 1-1 events in the data. The points outside the parallelogram in (a) are excluded from the fit. The distribution in (b) represents a profile histogram of the scatter plot for all points used in the maximum likelihood fit.

Before performing the fit, ten events lying in the tails of the impact parameter and acoplanarity distributions were removed (this corresponds to a 1.4% trim). This was done by defining a fit region represented by a parallelogram surrounding the core of the scatter plot in Fig. 7(a). In addition, in order to reproduce the distribution of impact parameter difference vs. acoplanarity as provided by Monte Carlo, the binning used in the maximum likelihood fit was chosen to be very fine in the middle of the scatter plot and progressively coarser towards the tails. In order to account for background, Monte Carlo events from the various sources of contamination were merged with the tau Monte Carlo sample used in the fit. The result of the fit for the tau lifetime using this technique is:

$$\tau_\tau = 298 \pm 13 \text{ fs}.$$

The error is statistical only. A comparison between the scatter plot in Fig. 7(a) and its counterpart from Monte Carlo yields a $\chi^2$ per degree of freedom of 1.2.

The largest source of systematic errors comes from the fitting procedure. We assign a systematic error of 1.6% due to the sensitivity of the measurement to the chosen bin size. This is due mainly to the tails of the impact parameter difference and acoplanarity distributions where the finite statistics in the Monte Carlo do not allow a very fine binning. In the Monte Carlo sample, we require a minimum of 5 events in each square bin in impact parameter difference and acoplanarity in order to compute the likelihood probability for that bin. When this is not satisfied, entries from neighboring bins are progressively combined until the minimum number of events is reached. We have varied the number of entries required from 1 to 10 and observe no change in the measured lifetime. We have also studied the effect of the size of the fit region which we have varied over a wide range. This lead to assigning a systematic error of 1.0%. Furthermore, we estimate a systematic error of 0.7% due to Monte Carlo statistics, determined from the spread in the measured lifetime as a function of the size of the analyzing Monte Carlo sample used in the maximum likelihood fit.

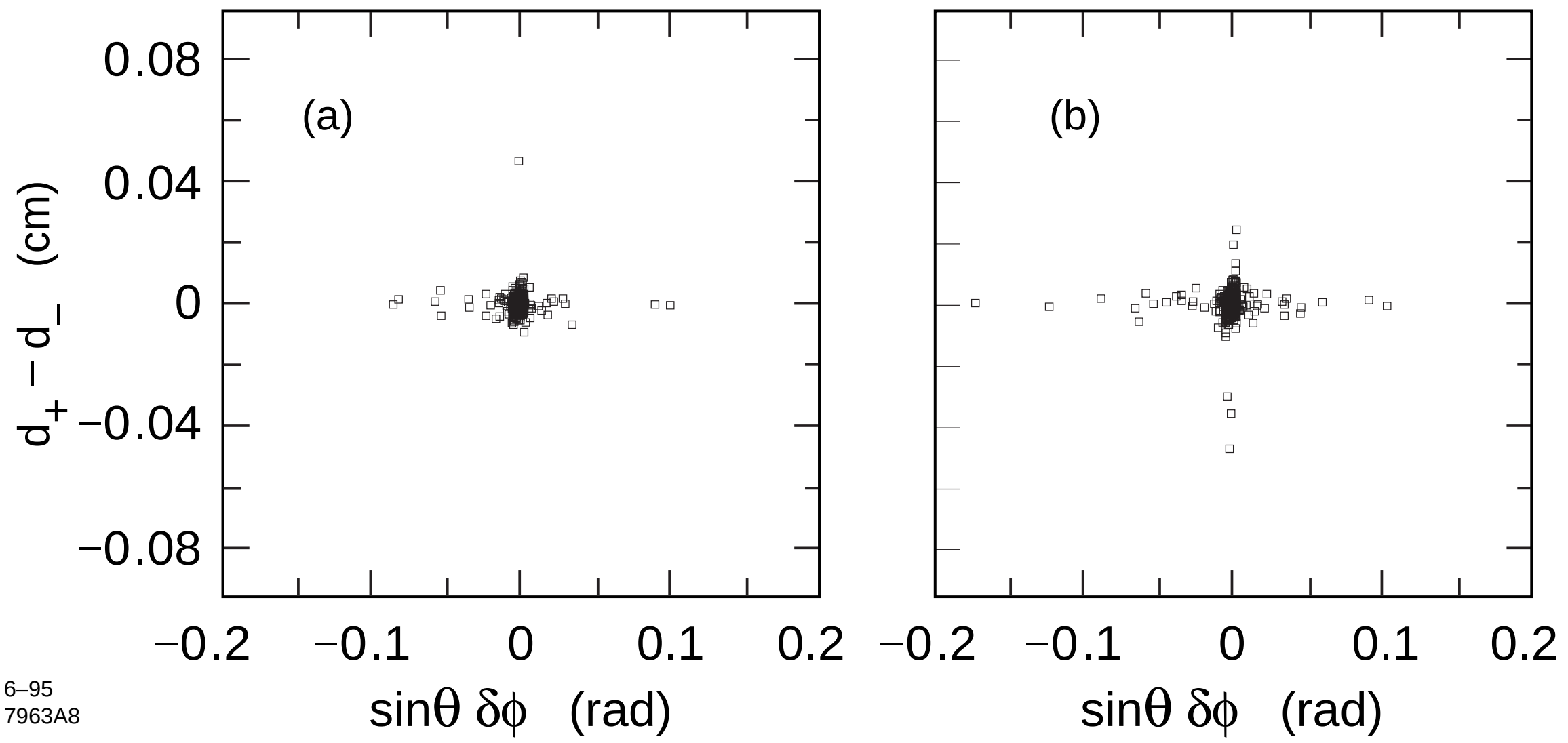

Figure 8: Impact parameter difference vs. acoplanarity scatter plots for (a) muon-pair and (b) wide-angle Bhabha events from the data.

Various analysis cuts were varied to study the effect of the event selection and the track quality criteria. We observe a 0.9% change in the measured lifetime when we vary the cuts that had the most effect at reducing the remaining background in the 1-1 sample, namely the cuts on the two-prong mass, the maximum track momentum, and $\cos\Theta_{miss}$. By varying the minimum track momentum and the minimum number of VXD and CDC hits on a track, we assign a systematic error of 0.5% due to track quality cuts. The sensitivity to the detector alignment was studied by dividing the data into bins in both the azimuth and polar angle; no significant effect was observed.

| Systematic effects | Error (%) |
|---|---|
| Binning | 1.6 |
| Fit range | 1.0 |
| Monte Carlo statistics | 0.7 |
| Event selection | 0.9 |
| Track quality cuts | 0.5 |
| Tau branching ratios | 0.2 |
| Initial and final state radiation | 0.3 |
| Beam energy and energy spread | 0.3 |
| Total | 2.3 |

Table 4: Systematic errors for the impact parameter difference method.

Similarly to the IP method, we assign a systematic uncertainty of 0.2% due to possible variations in the impact parameter distribution associated with each one-prong decay channel of the tau. Furthermore, we estimate a contribution of 0.3% from both initial- and final-state radiation and the uncertainty in the beam energy and the beam-energy spread.

Other systematic effects were investigated. The analysis was performed on data samples from different run periods; the results were all consistent. An important systematic check of the impact parameter difference technique was done by running on event samples with known zero lifetimes, namely Bhabha scattering and muon-pair final states. This is illustrated in Fig. 8 which shows the impact parameter difference versus acoplanarity for (a) muon-pair and (b) wide-angle Bhabha events, both taken from the data. The two distributions are flat as expected, and the majority of the events are clustered at the origin[17]. As a result, this method is relatively insensitive to background.

Replacing the data sample by several Monte Carlo data sets with the same number of events resulted in a spread in the measured lifetimes consistent with the statistical errors, while their mean value was very close to the input value in the Monte Carlo. This is a check that the method does not bear any systematic bias. This was further confirmed using several generator-level Monte Carlo samples with different lifetimes; the value of the tau lifetime with which the sample was generated was reproduced in each case.

The systematic errors are summarized in Table 4. Adding all contributions in quadrature leads to an overall systematic error of 2.3%. Thus, the result from the impact parameter difference method is:

$$\tau_\tau = 298 \pm 13 \ \pm 7 \text{ fs},$$

where the first error is statistical and the second is systematic.

# 5 Summary and Conclusions

We have measured the tau lifetime using three different techniques. The results from the three methods are consistent with one another. The decay length method is independent of the other two techniques since it uses a completely separate set of events. The impact parameter and impact parameter difference methods share about 70% of the events. However, because they make use of different information, they are not totally correlated. We have determined the correlation between the two techniques using Monte Carlo and following the procedure described in Ref. [18]. For the events that are common to the IP and IPD methods, we obtain a correlation factor of 53%. This results in an overall correlation of 37% between the two measurements. In this evaluation, common systematic effects are also taken into account.

A combined result of $\tau_\tau = 300 \pm 11(stat.) \pm 5(syst.)$ fs is derived from the IP and IPD measurements. With the inclusion of the DL measurement, we obtain:

$$\tau_\tau = 297 \pm 9(stat.) \pm 5(syst.) \text{ fs}.$$

This result is in agreement with the present world average value[19]. Our measurement is currently limited by the relatively small size of our data sample; we expect a significant improvement in the future as we anticipate a substantial increase in statistics, together with an associated decrease in the systematic error. A precise measurement of the tau lifetime at SLD, performed in a different environment relative to other experiments because of the small beam size and three-dimensional high-resolution vertexing, will represent an important contribution.

We wish to thank the SLC technical staff for their outstanding achievements in bringing micron-size beams into collision at the SLD with remarkable reliability.

## The SLD Collaboration

K. Abe,[29] I. Abt,[14] C.J. Ahn,[26] T. Akagi,[27] N.J. Allen,[4] W.W. Ash,[27]† D. Aston,[27] K.G. Baird,[25] C. Baltay,[33] H.R. Band,[32] M.B. Barakat,[33] G. Baranko,[10] O. Bardon,[16] T. Barklow,[27] A.O. Bazarko,[11] R. Ben-David,[33] A.C. Benvenuti,[2] T. Bienz,[27] G.M. Bilei,[22] D. Bisello,[21] G. Blaylock,[7] J.R. Bogart,[27] T. Bolton,[11] G.R. Bower,[27] J.E. Brau,[20] M. Breidenbach,[27] W.M. Bugg,[28] D. Burke,[27] T.H. Burnett,[31] P.N. Burrows,[16] W. Busza,[16] A. Calcaterra,[13] D.O. Caldwell,[6] D. Calloway,[27] B. Camanzi,[12] M. Carpinelli,[23] R. Cassell,[27] R. Castaldi,[23](a) A. Castro,[21] M. Cavalli-Sforza,[7] E. Church,[31] H.O. Cohn,[28] J.A. Coller,[3] V. Cook,[31] R. Cotton,[4] R.F. Cowan,[16] D.G. Coyne,[7] A. D'Oliveira,[8] C.J.S. Damerell,[24] M. Daoudi,[27] R. De Sangro,[13] P. De Simone,[13] R. Dell'Orso,[23] M. Dima,[9] P.Y.C. Du,[28] R. Dubois,[27] B.I. Eisenstein,[14] R. Elia,[27] E. Etzion,[4] D. Falciai,[22] M.J. Fero,[16] R. Frey,[20] K. Furuno,[20] T. Gillman,[24] G. Gladding,[14] S. Gonzalez,[16] G.D. Hallewell,[27] E.L. Hart,[28] Y. Hasegawa,[29] S. Hedges,[4] S.S. Hertzbach,[17] M.D. Hildreth,[27] J. Huber,[20] M.E. Huffer,[27] E.W. Hughes,[27] H. Hwang,[20] Y. Iwasaki,[29] D.J. Jackson,[24] P. Jacques,[25] J. Jaros,[27] A.S. Johnson,[3] J.R. Johnson,[32] R.A. Johnson,[8] T. Junk,[27] R. Kajikawa,[19] M. Kalelkar,[25] H. J. Kang,[26] I. Karliner,[14] H. Kawahara,[27] H.W. Kendall,[16] Y. Kim,[26] M.E. King,[27] R. King,[27] R.R. Kofler,[17] N.M. Krishna,[10] R.S. Kroeger,[18] J.F. Labs,[27] M. Langston,[20] A. Lath,[16] J.A. Lauber,[10] D.W.G. Leith,[27] M.X. Liu,[33] X. Liu,[7] M. Loreti,[21] A. Lu,[6] H.L. Lynch,[27] J. Ma,[31] G. Mancinelli,[22] S. Manly,[33] G. Mantovani,[22] T.W. Markiewicz,[27] T. Maruyama,[27] R. Massetti,[22] H. Masuda,[27] E. Mazzucato,[12] A.K. McKemey,[4] B.T. Meadows,[8] R. Messner,[27] P.M. Mockett,[31] K.C. Moffeit,[27] B. Mours,[27] G. Müller,[27] D. Muller,[27] T. Nagamine,[27] U. Nauenberg,[10] H. Neal,[27] M. Nussbaum,[8] Y. Ohnishi,[19] L.S. Osborne,[16] R.S. Panvini,[30] H. Park,[20] T.J. Pavel,[27] I. Peruzzi,[13](b) M. Piccolo,[13] L. Piemontese,[12] E. Pieroni,[23] K.T. Pitts,[20] R.J. Plano,[25] R. Prepost,[32] C.Y. Prescott,[27] G.D. Punkar,[27] J. Quigley,[16] B.N. Ratcliff,[27] T.W. Reeves,[30] J. Reidy,[18] P.E. Rensing,[27] L.S. Rochester,[27] J.E. Rothberg,[31] P.C. Rowson,[11] J.J. Russell,[27] O.H. Saxton,[27] S.F. Schaffner,[27] T. Schalk,[7] R.H. Schindler,[27] U. Schneekloth,[16] B.A. Schumm,[15] A. Seiden,[7] S. Sen,[33] V.V. Serbo,[32] M.H. Shaevitz,[11] J.T. Shank,[3] G. Shapiro,[15] S.L. Shapiro,[27] D.J. Sherden,[27] K.D. Shmakov,[28] C. Simopoulos,[27] N.B. Sinev,[20] S.R. Smith,[27] J.A. Snyder,[33] P. Stamer,[25] H. Steiner,[15] R. Steiner,[1] M.G. Strauss,[17] D. Su,[27] F. Suekane,[29] A. Sugiyama,[19] S. Suzuki,[19] M. Swartz,[27] A. Szumilo,[31] T. Takahashi,[27] F.E. Taylor,[16] E. Torrence,[16] J.D. Turk,[33] T. Usher,[27] J. Va'vra,[27] C. Vannini,[23] E. Vella,[27] J.P. Venuti,[30] R. Verdier,[16] P.G. Verdini,[23] S.R. Wagner,[27] A.P. Waite,[27] S.J. Watts,[4] A.W. Weidemann,[28] E.R. Weiss,[31] J.S. Whitaker,[3] S.L. White,[28] F.J. Wickens,[24] D.A. Williams,[7] D.C. Williams,[16] S.H. Williams,[27] S. Willocq,[33] R.J. Wilson,[9] W.J. Wisniewski,[5] M. Woods,[27] G.B. Word,[25] J. Wyss,[21] R.K. Yamamoto,[16] J.M. Yamartino,[16] X. Yang,[20] S.J. Yellin,[6] C.C. Young,[27] H. Yuta,[29] G. Zapalac,[32] R.W. Zdarko,[27] C. Zeitlin,[20] Z. Zhang,[16] and J. Zhou,[20]

[1]*Adelphi University, Garden City, New York 11530*
[2]*INFN Sezione di Bologna, I-40126 Bologna, Italy*
[3]*Boston University, Boston, Massachusetts 02215*

[4]*Brunel University, Uxbridge, Middlesex UB8 3PH, United Kingdom*
[5]*California Institute of Technology, Pasadena, California 91125*
[6]*University of California at Santa Barbara, Santa Barbara, California 93106*
[7]*University of California at Santa Cruz, Santa Cruz, California 95064*
[8]*University of Cincinnati, Cincinnati, Ohio 45221*
[9]*Colorado State University, Fort Collins, Colorado 80523*
[10]*University of Colorado, Boulder, Colorado 80309*
[11]*Columbia University, New York, New York 10027*
[12]*INFN Sezione di Ferrara and Università di Ferrara, I-44100 Ferrara, Italy*
[13]*INFN Lab. Nazionali di Frascati, I-00044 Frascati, Italy*
[14]*University of Illinois, Urbana, Illinois 61801*
[15]*Lawrence Berkeley Laboratory, University of California, Berkeley, California 94720*
[16]*Massachusetts Institute of Technology, Cambridge, Massachusetts 02139*
[17]*University of Massachusetts, Amherst, Massachusetts 01003*
[18]*University of Mississippi, University, Mississippi 38677*
[19]*Nagoya University, Chikusa-ku, Nagoya 464 Japan*
[20]*University of Oregon, Eugene, Oregon 97403*
[21]*INFN Sezione di Padova and Università di Padova, I-35100 Padova, Italy*
[22]*INFN Sezione di Perugia and Università di Perugia, I-06100 Perugia, Italy*
[23]*INFN Sezione di Pisa and Università di Pisa, I-56100 Pisa, Italy*
[25]*Rutgers University, Piscataway, New Jersey 08855*
[24]*Rutherford Appleton Laboratory, Chilton, Didcot, Oxon OX11 0QX United Kingdom*
[26]*Sogang University, Seoul, Korea*
[27]*Stanford Linear Accelerator Center, Stanford University, Stanford, California 94309*
[28]*University of Tennessee, Knoxville, Tennessee 37996*
[29]*Tohoku University, Sendai 980 Japan*
[30]*Vanderbilt University, Nashville, Tennessee 37235*
[31]*University of Washington, Seattle, Washington 98195*
[32]*University of Wisconsin, Madison, Wisconsin 53706*
[33]*Yale University, New Haven, Connecticut 06511*
[†]*Deceased*
[a]*Also at the Università di Genova*
[b]*Also at the Università di Perugia*